\documentclass[a4paper]{article}

\usepackage{INTERSPEECH2021}
\usepackage{graphicx}
\usepackage{multirow}
\usepackage[
separate-uncertainty = true,
multi-part-units = repeat
]{siunitx}
\usepackage{etoolbox}
\robustify\bfseries
\newrobustcmd\B{\DeclareFontSeriesDefault[rm]{bf}{b}\bfseries}  

\title{Rapping-Singing Voice Synthesis based on Phoneme-level Prosody Control}
\name{
	\begin{tabular}{c}
		Konstantinos Markopoulos$^{\star}$,
		Nikolaos Ellinas$^{\star}$,
		Alexandra Vioni$^{\star}$,
		Myrsini Christidou$^{\star}$, \\
		Panos Kakoulidis$^{\star}$,
		Georgios Vamvoukakis$^{\star}$,
		Georgia Maniati$^{\star}$,
		June Sig Sung$^{\dagger}$, \\
		Hyoungmin Park$^{\dagger}$,
		Pirros Tsiakoulis$^{\star}$,
		Aimilios Chalamandaris$^{\star}$
	\end{tabular}
}
\address{$^{\star}$ Innoetics, Samsung Electronics, Greece \\
	$^{\dagger}$ Mobile Communications Business, Samsung Electronics, Republic of Korea}
\email{\{k.markop, n.ellinas\}@samsung.com, \\
	\{a.vioni, m.christidou\}@partner.samsung.com, \\
	\{p.kakoulidis, g.vamvouk, g.maniati, js6.sung, hm94.park, p.tsiakoulis, aimilios.ch\}@samsung.com }

\begin{document}

\maketitle
\begin{abstract}
In this paper, a text-to-rapping/singing system is introduced, which can be adapted to any speaker's voice.
It utilizes a Tacotron-based multispeaker acoustic model trained on read-only speech data and which provides prosody control at the phoneme level. Dataset augmentation and additional prosody manipulation based on traditional DSP algorithms are also investigated.
The neural TTS model is fine-tuned to an unseen speaker's limited recordings, allowing rapping/singing synthesis with the target's speaker voice.
The detailed pipeline of the system is described, which includes the extraction of the target pitch and duration values from an a capella song and their conversion into target speaker's valid range of notes before synthesis.
An additional stage of prosodic manipulation of the output via WSOLA is also investigated for better matching the target duration values.
The synthesized utterances can be mixed with an instrumental accompaniment track to produce a complete song.
The proposed system is evaluated via subjective listening tests as well as in comparison to an available alternate system which also aims to produce synthetic singing voice from read-only training data.
Results show that the proposed approach can produce high quality rapping/singing voice with increased naturalness.
\end{abstract}

\section{Introduction}
With the recent development of neural text-to-speech (TTS), the task of singing voice synthesis (SVS) is gaining popularity, since it has become feasible to produce natural and expressive speech more effectively.
Before the development of deep neural network based synthesis, SVS systems were mainly based on unit selection technology  \cite{macon1997concatenation, hunt1996unit, gu2016singing, bonada2016expressive} or parametric TTS \cite{nakamura2014hmm, saino2006hmm}. 
During the last few years, with the establishment of neural TTS systems such as Tacotron \cite{wang2017tacotron}, it has become possible to investigate approaches like neural rapping and singing.

SVS is a complicated task. As in regular TTS, a large and powerful neural model that accurately predicts acoustic features must be designed and tuned. Additionally, singing information such as musical notes and rhythm must be accurately followed, in order to produce high quality samples.
For rapping, the procedure is the same, though the focus is less on musical notes, and more on pitch variation and rhythm, the latter translating into accurate phoneme durations.

\subsection{Related work}
Several approaches of rapping and singing voice synthesis have been proposed over the years. Early methods were based on unit concatenation \cite{macon1997concatenation, hunt1996unit, gu2016singing, bonada2016expressive} and statistical parametric synthesis \cite{nakamura2014hmm, saino2006hmm}. There was also an attempt focused on speech-to-rap voice conversion, based on a phase vocoder and beat tracking \cite{wu2014speechrap}. Nevertheless, such approaches had significant limitations and did not achieve high quality synthesized speech.
In recent years, many steps have been made towards high quality SVS, with the use of neural and deep learning methods. Hybrid, mixed and conditioned models have been introduced that advance further the SVS systems. 
Some notable approaches are the WaveNet variant architecture \cite{blaauw2017neural} used for parametric singing synthesis, WGANSing which is a pitch conditioned Generative Adversarial Network (GAN) \cite{chandna2019wgansing} and an adversarially trained, pitch conditioned sequence-to-sequence Korean singing model \cite{lee2019adversarially}.
Another GAN-based approach is unsupervised cross-domain singing voice conversion \cite{polyak2020ucdsvc}, which uses additional perceptual losses on its generator output. 
Mellotron \cite{valle2020mellotron}, a multi-speaker expressive voice synthesis model based on GST-Tacotron 2 \cite{shen2018natural}, also has SVS capabilities. 
Moreover, Jukebox \cite{dhariwal2020jukebox} generates singing voice with accompaniments, and UTACO \cite{angelini2020singing} consists of an attention-based sequence-to-sequence mechanism and a vocoder with dilated causal convolutions. 
Another approach was Durian-SC, a duration-informed, phoneme-to-acoustic features' alignment model with composite conditioning that allows SVS \cite{zhang2020duriansc}. 
The transformer architecture has also been employed to address the problem, as in DeepSinger \cite{ren2020deepsinger}, which employs separate encoders for its features, and HiFiSinger \cite{chen2020hifisinger}, which includes a FastSpeech-based \cite{ren2019fastspeech} model and multi-scale adversarial training approaches. 
Recently, ByteSing \cite{gu2021bytesing} was introduced, a Tacotron 2 model combined with a duration prediction model and uses linguistic along with musical embeddings.

\begin{figure*}[t]
	\centering
	\includegraphics[width=\linewidth]{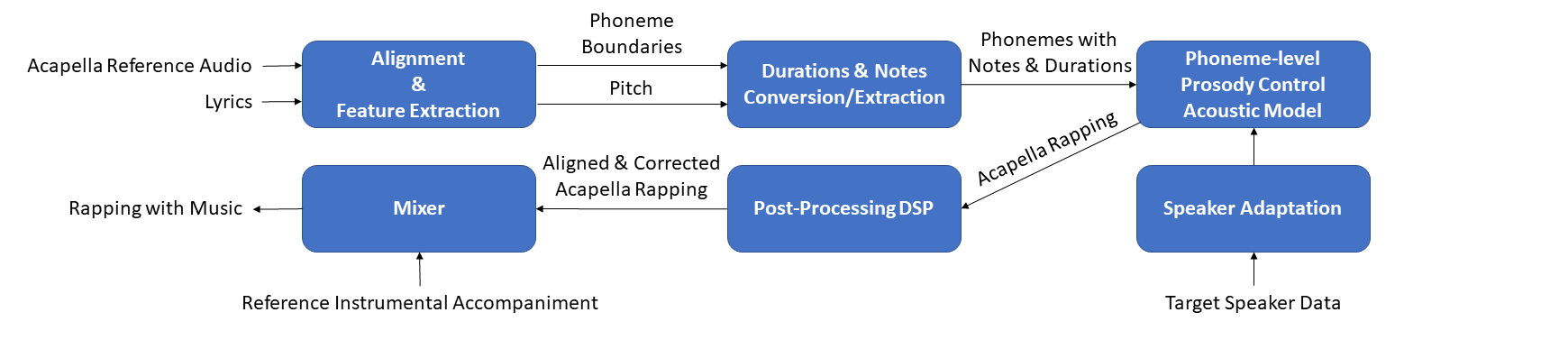}
	\caption{Proposed text-to-rapping/singing system}
	\label{fig:rapphoto1}
	\vspace{-0.6em}
\end{figure*}

\subsection{Proposed method}

In this paper, we propose a complete text-to-rapping and singing approach based on a system that, unlike other methods, relies solely on spoken data and can be adapted to an unseen target voice with very limited data.
Our method is based on a fine-grained prosody manipulation multispeaker TTS model presented in \cite{prosodycontrol2}.
Combined with augmentation of training data, our method can achieve phoneme-level prosody manipulation (F0 and duration), which allows us to generate rapping and singing synthesized speech.

We train a multispeaker multilingual TTS model on internal spoken data, in US English and Korean languages, and show that the prosody control capabilities at the phoneme-level musical note and duration are effective in all training speakers.
For the case of speaker adaptation with very limited data, we resume the model training on 2 speakers of both languages with only about 11 minutes of spoken data, and show that the resulting models can also maintain the same prosody control capabilities.
Extracting the desired prosody information from a given a capella song and presenting it as input to our model enables us to synthesize a capella utterances that closely follow the reference, allowing rapping/singing synthesis with the voice of every speaker included in the training set.

We present the proposed approach, from the stage of phoneme alignment and F0 extraction and discretization, up to the final post-processing steps required to produce the actual song with the voice of the selected speaker.
This includes several processing steps and modules as well as a fine-tuning digital signal processing (DSP) stage based on synchronous overlap-add algorithms \cite{charpentier1986diphone, verhelst1993overlap}, in order to achieve exact alignment of the synthetic speech to the target song music track.
Finally, our system is evaluated via crowd-sourced listening tests against the ground truth samples, as well as against samples from Mellotron, a widely accepted state-of-the-art SVS system.
Objective evaluation results are also presented, showing that our model follows the desired prosody patterns appropriately.

\section{Method}

\subsection{Overview}

Our proposed text-to-rapping/singing system aims at producing a song with the voice of a target speaker, based only on the lyrics and an a capella version of the song by its original singer.
A block diagram of the system is shown in Figure~\ref{fig:rapphoto1} and consists of four main parts.

The first part involves the preprocessing of the lyrics by a front-end module in order to obtain the phonemes, as well as the extraction of the required prosodic features from the reference audio, i.e. phoneme durations and F0 (Section~\ref{sec:alignment}).
Second, to avoid extreme F0 values which can hinder the output quality, F0 contours are converted so as to lie within the range of the target speaker (Section~\ref{sec:notesection}).
Third, the converted F0 values along with the phoneme durations are discretized and used by the TTS model, which produces the synthesized song by controlling the prosody at the phoneme-level (Section~\ref{sec:acoustic}).
The fourth step is post-processing, including the introduction of a DSP module, which is necessary for accurate time alignment of the produced song with the original time specification, if the instrumental accompaniment track needs to be combined.
This task can be performed by a mixer which combines the synthesized a capella utterances with the music track (Section~\ref{sec:postprocessing}).
A detailed analysis of each step is given in the following sections.

\subsection{Alignment \& feature extraction}
\label{sec:alignment}

A capella songs are used as reference since the acoustic model is trained only on neutral style spoken data which do not contain music tracks.
Initially, phoneme alignments are automatically extracted from the song using an HMM monophone acoustic model trained using flat start initialization \cite{raptis2016expressive}.
As the forced-alignment model is trained only on spoken utterances and may not produce precise phoneme boundaries, a manual correction of the alignments was performed.
Most corrections were attributed to accommodate for long vowel durations in the singing data, which are not usually encountered in spoken speech.

F0 contours are calculated using the algorithm included in the Praat toolkit \cite{boersma2006praat}.
Interpolation is applied on the unvoiced regions in order to avoid zero values and the final contour is smoothed.

\subsection{Note conversion/extraction}
\label{sec:notesection}

The acoustic model is trained to control the prosody within the range of each speaker.
When a reference song is used as input to the proposed system, the F0 contour of the singing speaker may have different range than that of the target speaker.
This is due to the fact that the song originates from a different speaker with diverse source characteristics or gender. Also, the singing speech itself may vary in extreme F0 values both in the lower and higher end between two speakers.

We use Eq~(\ref{eq:notes3}) in order to transpose the F0 contour of the reference speaker to match the range of the target speaker:

\begin{equation}
	f_{target}=\frac{median(f_{speaker})}{median(f_{ref})}\cdot f_{ref}
	\label{eq:notes3}
\end{equation}
where $f_{speaker}$ and $f_{ref}$ represents all the F0 values of the target and reference speaker's utterances respectively.

Two approaches were examined on F0 transposition.
The first approach takes one single (global) pitch value for the reference speaker, which is the median F0 value, while the second recalculates the median F0 value for every verse.
Early results demonstrated that the former approach works better, producing more stable samples, and avoids the discontinuities in octaves that may occur due to the recalculation of the F0.

After the conversion, the average phoneme-level F0 is calculated by using the previously extracted alignments.
The corresponding musical note and octave can be extracted by using formulas (\ref{eq:notes1}) and (\ref{eq:notes2}):

\begin{equation}
h=\left\lfloor 12\cdot\log_2\frac{f_{target}}{440} \right\rceil+57
\label{eq:notes1}
\end{equation}
\begin{equation}
\quad\mathrm{}\quad
octave=\left\lfloor \frac{h}{12} \right\rfloor
\quad\mathrm{and}\quad
note=\left(h\bmod\ 12\right)
\label{eq:notes2}
\end{equation}
where $h$ represents the distance in semitones from the note $C_0$.

\subsection{Acoustic model}
\label{sec:acoustic}

For controlling prosody, we use an acoustic model based on previous work \cite{prosodycontrol}.
This method uses unsupervised clustering on phoneme-level F0 and duration values in order to extract a sequence of discrete learnable prosodic labels for each utterance, which is then used to condition the decoder of a Tacotron-based acoustic model \cite{wang2017tacotron,shen2018natural} in parallel to the phoneme sequence.
The model is also augmented with a secondary Mixture-of-Logistics (MoL) attention module \cite{lpctron} which operates on the prosodic sequence only, aiming to disentangle the phonetic and prosodic content.
The final model is able to control the prosody at the phoneme level both for F0 and duration while maintaining high speech quality.

In \cite{prosodycontrol}, it is shown that this model is also effective at representing musical notes instead of F0 values that are derived from K-means unsupervised clustering.
In the current work, we follow the procedure described in Section~\ref{sec:notesection} to assign the notes and octaves to separate learnable embeddings, so that any possible combinations are modeled appropriately, even non-existing ones in the training set.
This global representation of musical notes is also speaker-independent, being suitable for our multi-speaker setup described below.

For the duration labels, we follow an improved and more stable method than K-means, as in \cite{prosodycontrol2}.
The values are sorted in ascending order and grouped into a desired number of intervals, so that an equal number of samples is contained in each interval.
This alleviates the problem of voice quality deterioration in extreme values which are not common in the training set, while slightly decreasing the duration control range. In this work, in order to investigate the effect of the number of labels involved, we have experimented with two different setups, one with 15 and another with 30 duration labels.

The training setup follows a multi-speaker/multi-lingual scenario, similar to the parallel work done in \cite{prosodycontrol2}.
A multitude of speakers of both genders and 2 languages is used instead of a single one, in order to capture as many prosody patterns as possible, which will help increase the range of the model in both F0 and duration.
That way each speaker has the capability to rap/sing a wider variety of songs.
Augmentation is also employed both in F0 and duration by applying pitch shifting and tempo alteration respectively, further increasing the quantity and range of the training data.
This setup also allows us to perform speaker adaptation using limited data from unseen speakers and enable rapping/singing for the adaptation speaker in both languages.
This process becomes easier as the model does not have unseen values due to the global musical note representation for the F0 and the duration intervals which are derived from all speakers and are common throughout the training procedure.

The different speakers are assigned a learnable speaker embedding, which also conditions the decoder at each step.
We also use a linear adversarial speaker classifier on the phoneme encoder outputs, in order to make them speaker independent, as well as a residual variational encoder which captures other latent factors of the recordings \cite{Zhang2019}.
This method is shown to improve naturalness and stability by simply using a zero vector during inference, which is essentially the prior mean.
The multilingual setup does not include language embeddings, but simply considering every phone for each language with a different label, increasing the total number of phones to the sum of each language phoneset.
For the speaker adaptation, we found that freezing the weights of the phoneme encoder, prosody encoder and attention modules yields better results.
We can account this to the fact that the model has already learned rich representations which must not be forgotten in the adaptation stage by the target speaker's limited data.

\subsection{Post-processing}
\label{sec:postprocessing}

The multi-speaker/multi-lingual prosody control model produces a capella utterances which are derived between silence tokens from the original lyrics.
These utterances must be concatenated in the time domain in order to obtain the full verse of the song.
Additionally, the duration values are discrete in our model, resulting in some form of quantization, hence the final durations may not exactly match the original song.
The accurate matching of the synthesized song with the original, requires lengthening or shortening of speech, which in our case is performed utilizing time-domain DSP methods.

This stage is based on two main algorithms: WSOLA \cite{verhelst1993overlap} and PSOLA \cite{laprie1998automatic}.
These algorithms are both able to modify the duration of each phoneme without affecting the pitch and resynthesize the original audio using the overlap-add technique.
In early listening experiments, we found that WSOLA produced slightly better results in almost every comparison, so we included this method in the evaluations that follow.
A Mixer element can also be included for producing the final song, mixing the time-aligned processed synthetic a capella utterance with the respective music track.
Mixing more than one voices is also possible, providing a multi-speaker song.

\section{Experiments \& results}
In this section, we describe our experimental setup, along with the method followed to objectively and subjectively evaluate the proposed system. The evaluation results are then discussed.

\subsection{Experimental setup}

\begin{table}[!t]
	\caption{Multilingual multi-speaker dataset for text-to-singing/rapping model training. `tr' and `ad' refer to training and adaptation speakers respectively, while `f' and `m' refer to gender.}
	\label{table:marky}
	\centering
	\begin{tabular}{cccc}
		\toprule
		\textbf{Speaker} &  \textbf{Language} &  \textbf{Rec. Hours} &  \textbf{Utterances}  \\
		\midrule
		us\_tr\_f1  & en-us & 41.21 & 36185 \\
		
		us\_tr\_f2  & en-us & 38.88 & 45841 \\
		
		us\_tr\_m1  & en-us & 36.82 & 40442 \\
		
		ko\_tr\_f1  & ko & 51.37 & 40503 \\
		
		ko\_tr\_f2  & ko & 54.29 & 29289 \\
		
		us\_ad\_m1  & en-us & 0.19 & 165 \\
		
		ko\_ad\_m1  & ko & 0.18 & 149 \\
		\bottomrule
	\end{tabular}
	\vspace{-10pt}
\end{table}

The acoustic model is trained with an internal multilingual multi-speaker dataset containing 222 hours of speech in both US English (en-us) and Korean (ko) from 1 male and 4 female speakers. The recorded dataset is a general TTS corpus of neutral speaking style.
For speaker adaptation, we use about 11 minutes of recordings from an unseen male speaker from each language.
The adaptation utterances are selected with a corpus selection process described in \cite{Chalamandaris2009} which ensures maximum phonetic coverage for the required amount of recordings.
Details regarding the data used for training and adaptation are presented in Table \ref{table:marky}.
The augmentations applied to the original training data, similar to \cite{prosodycontrol2}, include increasing and decreasing the F0 by 2, 4 or 6 semitones and the speaking rate from 70\% - 130\%. The final training set size after the augmentations is $415$ hours.

All audio data was resampled to 24 kHz.
The acoustic features were extracted in order to match the modified LPCNet Vocoder \cite{vipperla2020bunched} and consist of 20 Bark-scale cepstral coefficients, the pitch period and pitch correlation.
The phoneme encoder maps the input phoneme sequence into 256 dimensional embeddings and further applies a CBHG module.
In the prosody encoder, prosodic labels are mapped into 64 dimensional embeddings.
These are processed by a single 128-dimensional feed-forward Pre-Net with ReLU activation and a bidirectional Gated Recurrent Unit (GRU) layer with 128 dimensions in each direction.
The decoder contains 3 recurrent layers, a 256-dimensional attention GRU and two 512-dimensional residual LSTMs.
The attention modules used have a mixture of 5 logistic distributions and 256-dimensional feed-forward layers.
Dropout regularization \cite{srivastava2014dropout} of rate 0.5 is applied on all Pre-Net and Post-Net layers and Zoneout \cite{krueger2016zoneout} of rate 0.1 is applied on LSTM layers.
We use the Adam optimizer \cite{kingma2015adam} for training the network parameters with batch size 32. The learning rate is initially $10^{-3}$ and decays linearly to $3 \cdot 10^{-5}$ after $100,000$ iterations. We also apply L2 regularization with factor $10^{-6}$.
Speaker adaptation involves further training of the prosody model for 5K iterations, with frozen weights in the phoneme encoder, prosody encoder and attention modules.

\subsection{Objective evaluation}

\begin{table}[!t]
	\caption{Average semitone pitch \& duration (ms) score for objective evaluation }
	\vspace{-5pt}
	\label{table:objj}
	\centering
	\begin{tabular}{llcccc}
		\toprule
		& & \textbf{Training} & \textbf{Adaptation}  \\
		& & \textbf{Speakers} & \textbf{Speakers}  \\
		\midrule
		\multirow{3}{6em}{\textbf{Pitch Error (semitones)}} & 15-class & 0.85 $\pm$ 0.22  & 0.86 $\pm$ 0.23\\
		& 30-class & 0.93 $\pm$ 0.28  & 0.93 $\pm$ 0.21 \\
		& Mellotron & \multicolumn{2}{c}{0.51 $\pm$ 0.15} \\
		\midrule
		\multirow{3}{6em}{\textbf{Duration Error (msec)}} & 15-class & 27 $\pm$ 9  & 29 $\pm$ 8 \\
		& 30-class  & 29 $\pm$ 10  & 28 $\pm$ 7  \\
		& Mellotron & \multicolumn{2}{c}{25 $\pm$ 9} \\
		\bottomrule
	\end{tabular}
	\vspace{-10pt}
\end{table}

Objective evaluation of the proposed systems in terms of a capella song synthesis was attempted. Clips from 4 rap songs and 4 songs, all sung a capella in English, were used as ground truth for this evaluation\footnote{The reader is encouraged to listen to the audio samples at: https://innoetics.github.io/publications/rappotron/index.html}. All the proposed models along with Mellotron \cite{valle2020mellotron} were evaluated against the ground truth a capella songs. The samples for Mellotron were produced by Mellotron's latest GitHub repository along with a pretrained Mellotron model and respective WaveGlow model based on LibriTTS dataset. A random female voice from the model was selected for inference of the audio stimuli.

\begin{figure}[h]
	\centering
	\includegraphics[width=0.97\linewidth]{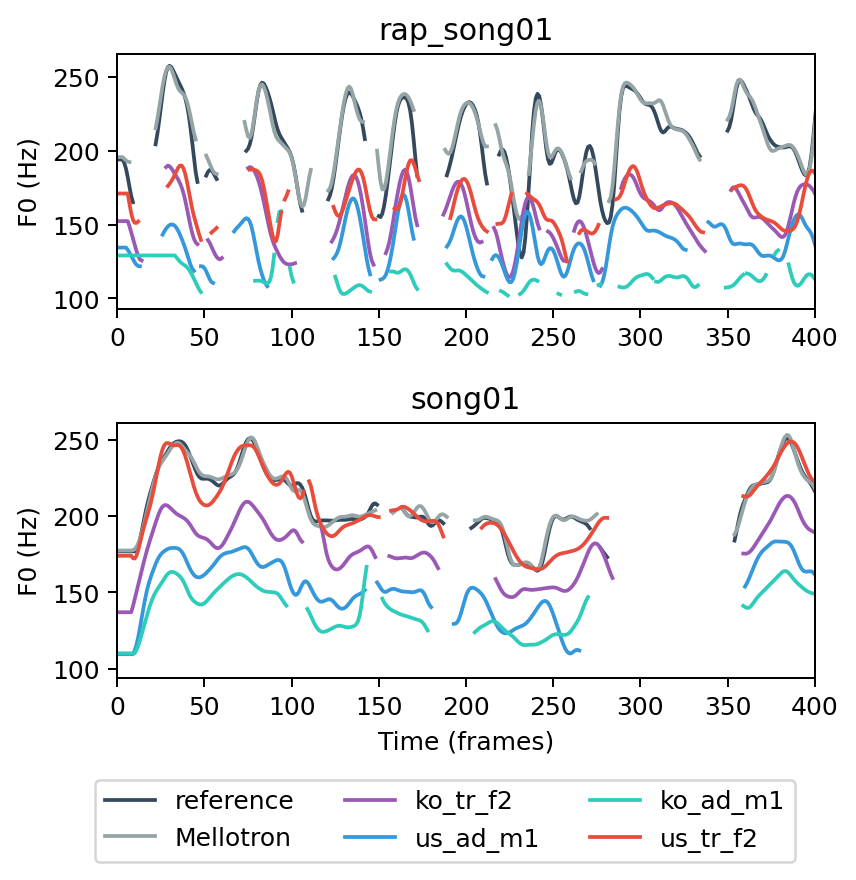}
	\caption{Pitch contours of the original song, Mellotron and proposed models.}
	\label{fig:obj_eval1}	
	\vspace{-10pt}
\end{figure}

\begin{figure}[h]
	\centering
	\includegraphics[width=0.97\linewidth]{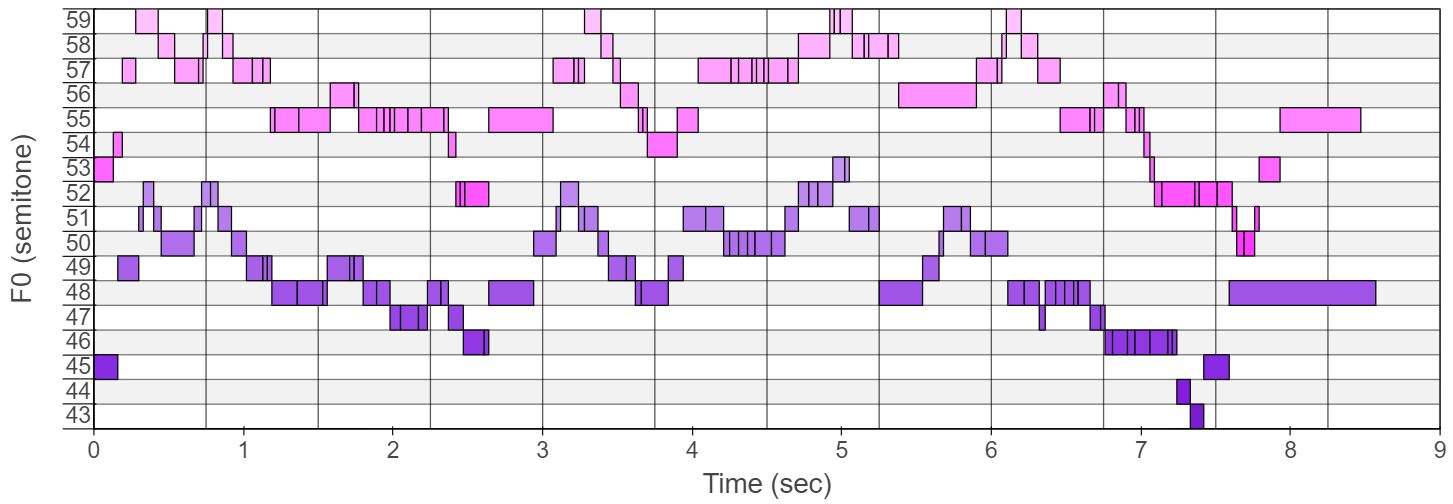}
	\caption{Pitch contours of the original and synthesized song in a MIDI representation (pink and purple values respectively). The MIDI values are calculated on a per phoneme basis.}
	\label{fig:obj_eval2}
	\vspace{-5pt}
\end{figure}

In Figure~\ref{fig:obj_eval1}, the pitch contours of 2 a capella reference songs are presented, along with the pitch contours of the audio stimuli inferred by the systems under comparison. 
One can notice that the synthesized pitch contours closely follow the reference ones, while they are shifted appropriately in order to match the F0 mean value of the inference speaker of each model.
In Figure~\ref{fig:obj_eval2}, where the F0 contours of a reference and synthesized clip are illustrated in a MIDI-like graph, a MIDI value is produced for each phoneme, and the trend of the synthesized melody following in parallel the ground truth one is obvious. 

\begin{table*}[!t]
	\footnotesize
	\caption{Mean Opinion Score (MOS) evaluation results with 95\% confidence interval}
	\vspace{-5pt}
	\label{table:mos}
	\centering
	\begin{tabular}{@{\hspace*{3mm}} l @{\hspace*{0mm}} S @{\hspace*{0mm}} S @{\hspace*{3mm}} S @{\hspace*{0mm}} S @{\hspace*{3mm}} S @{\hspace*{0mm}} S @{\hspace*{3mm}} S @{\hspace*{0mm}} S @{\hspace*{3mm}} S @{\hspace*{0mm}} S @{\hspace*{3mm}} S @{\hspace*{0mm}} S @{\hspace*{3mm}} S @{\hspace*{0mm}} S @{\hspace*{3mm}} S @{\hspace*{0mm}} S @{\hspace*{3mm}}}
		
		\toprule
		& \multicolumn{8}{c}{\textbf{30 duration labels}} & \multicolumn{8}{c}{\textbf{15 duration labels}} \\
		
		\midrule
		& \multicolumn{4}{c}{\textbf{rap\_songs}} & \multicolumn{4}{c}{\textbf{songs}} & \multicolumn{4}{c}{\textbf{rap\_songs}} & \multicolumn{4}{c}{\textbf{songs}} \\
		
		\midrule
		\textbf{setup} & \multicolumn{2}{c}{\textbf{plain}} & \multicolumn{2}{c}{\textbf{w/ post-proc}} & \multicolumn{2}{c}{\textbf{plain}} & \multicolumn{2}{c}{\textbf{w/ post-proc}} & \multicolumn{2}{c}{\textbf{plain}} & \multicolumn{2}{c}{\textbf{w/ post-proc}} & \multicolumn{2}{c}{\textbf{plain}} & \multicolumn{2}{c}{\textbf{w/ post-proc}} \\

		\midrule
		us\_tr\_f1 & 3.60 & {\scriptsize${\pm}$0.36} & 3.65  & {\scriptsize${\pm}$0.30} & 3.36 & {\scriptsize${\pm}$0.33} & 3.71  & {\scriptsize${\pm}$0.33} & 3.65 & {\scriptsize${\pm}$0.32} & 3.72 & {\scriptsize${\pm}$0.35} & 3.69 & {\scriptsize${\pm}$0.33} & 3.69 & {\scriptsize${\pm}$0.29} \\
		us\_tr\_f2 & 3.60 & {\scriptsize${\pm}$0.30} & 3.86 & {\scriptsize${\pm}$0.29} & 3.38 & {\scriptsize${\pm}$0.38} & 3.50 & {\scriptsize${\pm}$0.38} & 3.53 & {\scriptsize${\pm}$0.31} & 3.74 & {\scriptsize${\pm}$0.35} & 3.81 & {\scriptsize${\pm}$0.29} & 3.55 & {\scriptsize${\pm}$0.37} \\
		us\_tr\_m1 & 3.60 & {\scriptsize${\pm}$0.33} & 4.00 & {\scriptsize${\pm}$0.29} & 3.64 & {\scriptsize${\pm}$0.32} & 3.60 & {\scriptsize${\pm}$0.36} & 3.51 & {\scriptsize${\pm}$0.27} & 3.56 & {\scriptsize${\pm}$0.32} & 3.64 & {\scriptsize${\pm}$0.34} & 3.62 & {\scriptsize${\pm}$0.29} \\
		ko\_tr\_f1 & 3.35 & {\scriptsize${\pm}$0.33} & 3.70 & {\scriptsize${\pm}$0.36} & 3.17 & {\scriptsize${\pm}$0.34} & 3.33 & {\scriptsize${\pm}$0.41} & 3.44 & {\scriptsize${\pm}$0.37} & 3.35 & {\scriptsize${\pm}$0.34} & 3.81 & {\scriptsize${\pm}$0.32} & 3.57& {\scriptsize${\pm}$0.31} \\
		ko\_tr\_f2 & 3.37 & {\scriptsize${\pm}$0.34} & 3.42 & {\scriptsize${\pm}$0.35} & 3.45 & {\scriptsize${\pm}$0.39} & 3.33 & {\scriptsize${\pm}$0.37} & 3.40 & {\scriptsize${\pm}$0.32} & 3.40 & {\scriptsize${\pm}$0.34} & 3.69 & {\scriptsize${\pm}$0.35} & 3.81 & {\scriptsize${\pm}$0.32} \\
		us\_ad\_m1 & 3.67 & {\scriptsize${\pm}$0.35} & 3.74 & {\scriptsize${\pm}$0.29} & 3.74 & {\scriptsize${\pm}$0.35} & 3.33 & {\scriptsize${\pm}$0.37} & 3.79 & {\scriptsize${\pm}$0.31} & 3.51 & {\scriptsize${\pm}$0.35} & 3.57 & {\scriptsize${\pm}$0.34} & 3.33 & {\scriptsize${\pm}$0.39} \\
		ko\_ad\_m1 & 3.42 & {\scriptsize${\pm}$0.39} & 3.47 & {\scriptsize${\pm}$0.39} & 3.26 & {\scriptsize${\pm}$0.42} & 3.26 & {\scriptsize${\pm}$0.40} & 3.63 & {\scriptsize${\pm}$0.36} & 3.51 & {\scriptsize${\pm}$0.32} & 3.26 & {\scriptsize${\pm}$0.38} & 3.43 & {\scriptsize${\pm}$0.38} \\
		\midrule
		\textbf{Total} & \bfseries{3.52} &\B {\scriptsize${\pm}$0.34} & \bfseries{3.69} &\B {\scriptsize${\pm}$0.33} & \bfseries{3.43} &\B {\scriptsize${\pm}$0.36} & \bfseries{3.44} &\B {\scriptsize${\pm}$0.37} & \bfseries{3.56} &\B {\scriptsize${\pm}$0.32} & \bfseries{3.54} &\B {\scriptsize${\pm}$0.33} & \bfseries{3.64} &\B {\scriptsize${\pm}$0.33} & \bfseries{3.57} &\B {\scriptsize${\pm}$0.33}\\
		\midrule
		Mellotron & 3.33 & {\scriptsize${\pm}$0.38} & & & 3.17 & {\scriptsize${\pm}$0.49} & & & & & & & & & & \\
		\bottomrule        

	\end{tabular}
	\vspace{-10pt}
\end{table*}

As an objective metric for measuring the accuracy of our approach, we calculated the average distance of F0 and duration between the reference and the synthesized audio on a per phoneme basis. F0 values of the reference audio were transposed again, so as to target the speaker's mean F0 value before calculation of the distance. As depicted in Table \ref{table:objj}, there is no significant difference in these metrics between speakers seen in the training set and those used only for adaptation, neither on F0 nor on duration values. This observation is valid for both 15 and 30 duration label models. Mellotron seems to achieve an F0 contour closer to the reference, as was also illustrated in Figure~\ref{fig:obj_eval1}.

Overall, our method manages to reproduce a capella singing with accurate pitch and duration phoneme values especially in cases where the respective a capella reference audio is straightforward, without notably long durations or extreme low/high notes. In the latter cases, we noticed that our approach did not accurately produce phonemes with the target F0 or duration values, leading either to attention failures or duration mismatches with the reference audio.
 


\subsection{Subjective evaluation and discussion}

A subjective evaluation was carried out via a formal listening test. Two clips from songs and two clips from rap songs, all in English a capella, were used as ground truth, and the respective audio stimuli produced by our system and Mellotron were rated.
In the framework of this subjective evaluation, aside from the overall quality, we also opted to assess: a) the effect of the number of duration labels per phoneme, and b) the effect of the post-processing stage for duration matching to the reference audio via WSOLA (as described in Section~\ref{sec:postprocessing}). By combining the aforementioned parameters we produced all 4 possible models for each speaker described in Table~\ref{table:marky}.
In total, 60 listeners (Amazon Mechanical Turkers) participated in the listening test, rating each synthetic stimulus on a 5-point Likert scale on both melody and intelligibility, with 1 indicating ``Totally off-tune or wrong lyrics" and 5 indicating ``Exactly same melody and lyrics".


The average MOS and 95\% confidence interval for each voice model versus song type (rap song and song) and post-processing are presented in Table \ref{table:mos}.
The results show that our approach provides satisfactory output, equally for both rapping and singing, even if the latter is considered as a more complex task. 
Overall, post-processing for matching the word boundaries between ground-truth and audio stimuli does not seem to provide any consistent and robust improvement. Although there is no statistical significance in the pairwise differences observed between models, an improvement tendency is observed when post-processing is used for rap songs and the 30-duration-labeled model.
This post-processing stage remains necessary in order to align the generated songs with the music track at the final mixing stage, in case of music accompaniment of the synthesized a cappella voice. 
The 15-class duration labeling yields an improvement tendency for singing, which is most probably attributed to the fact that fewer but more populated classes in training entail better learning for our model.

A closer inspection of the results leads us to the conclusion that adapted speakers achieve similar quality results with the speakers who are seen during training. 
This is prominent especially for English, where the adapted voice performs equally well to the rest of the English training set voices, while, at the same time, the Korean adapted voice follows closely the performance of the Korean training voices.
Such similar MOS ratings confirm our hypothesis that our approach is robust for limited speaker data scenarios.
As far as the per speaker and language performance is concerned, Korean voice models have received lower MOS scores than the English ones. This is most probably due to the fact that the reference songs we evaluated are exclusively in English. Our informal evaluation showed that these non-native voice models mainly suffered from lower intelligibility or lack of naturalness, as they did not bear native English accent and thus sounded more artificial when synthesizing English a capella songs.

Our informal listening evaluation of the samples showed that Mellotron output is melodic but often bears intelligibility issues or audio artifacts. This observation may justify why Mellotron scored lower compared to our system in the formal subjective evaluation (Table~\ref{table:mos}).
Nevertheless, Mellotron outperformed our approach in a test song where vibrato singing was prominent, a voice characteristic that Mellotron can capture and transfer well, in contrast to our approach where only F0 and duration values per phoneme are provided.

\section{Conclusions}
In this paper, we presented an approach for producing high-quality singing and rapping synthesis from a Tacotron-based fine-grained prosody-control voice model trained solely on read data. Even though its results do not match the output of an SVS system trained on singing data, our approach achieves satisfactory results in both singing and rapping.
Experiments showed that equally good singing synthesis can be achieved for limited-data target voices via adaptation.
It is worth-noting that our system, similarly to other systems based on spoken-only data, suffers limitations in its ability to produce too long and extremely low or high-pitched sounds. In other words, although it can provide satisfactory results for rapping and simple songs, it may fail to produce adequate singing for more challenging songs with wide variations in both note values and duration. 
Another native limitation to our current approach is the lack of ability to transfer micro-prosodic characteristics into the synthetic output, such as vibrato or tremolo. Mellotron was shown to better imitate micro-prosodic singing voice qualities with the help of a more complex model and attention mechanism. 
Further research on controlling singing voice characteristics, such as loudness and vibrato, as well as on including singing data in the training process is required. 
Moreover, we plan to investigate ways for automating parts of the proposed process, such as the alignment optimization of the target singing data, so as to eliminate manual effort in our method. 
\bibliographystyle{IEEEtran}

\bibliography{refs}

\begin{thebibliography}{10}
\providecommand{\url}[1]{#1}
\csname url@samestyle\endcsname
\providecommand{\newblock}{\relax}
\providecommand{\bibinfo}[2]{#2}
\providecommand{\BIBentrySTDinterwordspacing}{\spaceskip=0pt\relax}
\providecommand{\BIBentryALTinterwordstretchfactor}{4}
\providecommand{\BIBentryALTinterwordspacing}{\spaceskip=\fontdimen2\font plus
\BIBentryALTinterwordstretchfactor\fontdimen3\font minus
  \fontdimen4\font\relax}
\providecommand{\BIBforeignlanguage}[2]{{%
\expandafter\ifx\csname l@#1\endcsname\relax
\typeout{** WARNING: IEEEtran.bst: No hyphenation pattern has been}%
\typeout{** loaded for the language `#1'. Using the pattern for}%
\typeout{** the default language instead.}%
\else
\language=\csname l@#1\endcsname
\fi
#2}}
\providecommand{\BIBdecl}{\relax}
\BIBdecl

\bibitem{macon1997concatenation}
M.~Macon, L.~Jensen-Link, E.~B. George, J.~Oliverio, and M.~Clements,
  ``Concatenation-based midi-to-singing voice synthesis,'' in \emph{Audio
  Engineering Society Convention 103}.\hskip 1em plus 0.5em minus 0.4em\relax
  Audio Engineering Society, 1997.

\bibitem{hunt1996unit}
A.~J. Hunt and A.~W. Black, ``Unit selection in a concatenative speech
  synthesis system using a large speech database,'' in \emph{1996 IEEE
  International Conference on Acoustics, Speech, and Signal Processing
  Conference Proceedings}, vol.~1.\hskip 1em plus 0.5em minus 0.4em\relax IEEE,
  1996, pp. 373--376.

\bibitem{gu2016singing}
H.-Y. Gu and J.-K. He, ``Singing-voice synthesis using demi-syllable unit
  selection,'' in \emph{2016 International Conference on Machine Learning and
  Cybernetics (ICMLC)}, vol.~2.\hskip 1em plus 0.5em minus 0.4em\relax IEEE,
  2016, pp. 654--659.

\bibitem{bonada2016expressive}
J.~Bonada, M.~Umbert~Morist, and M.~Blaauw, ``Expressive singing synthesis
  based on unit selection for the singing synthesis challenge 2016,''
  \emph{Morgan N, editor. Interspeech 2016; 2016 Sep 8-12; San Francisco, CA.
  ISCA; 2016. p. 1230-4.}, 2016.

\bibitem{nakamura2014hmm}
K.~Nakamura, K.~Oura, Y.~Nankaku, and K.~Tokuda, ``Hmm-based singing voice
  synthesis and its application to japanese and english,'' in \emph{2014 IEEE
  International Conference on Acoustics, Speech and Signal Processing
  (ICASSP)}.\hskip 1em plus 0.5em minus 0.4em\relax IEEE, 2014, pp. 265--269.

\bibitem{saino2006hmm}
K.~Saino, H.~Zen, Y.~Nankaku, A.~Lee, and K.~Tokuda, ``An hmm-based singing
  voice synthesis system,'' in \emph{Ninth International Conference on Spoken
  Language Processing}, 2006.

\bibitem{wang2017tacotron}
W.~et~al, ``Tacotron: Towards end-to-end speech synthesis,'' in \emph{Proc.
  Interspeech}, 2017.

\bibitem{wu2014speechrap}
M.~{Wu}, C.~{Lu}, and J.~R. {Jang}, ``Automatic conversion from speech to rap
  music,'' in \emph{2014 International Conference on Electrical Engineering and
  Computer Science (ICEECS)}, 2014, pp. 245--250.

\bibitem{blaauw2017neural}
M.~Blaauw and J.~Bonada, ``A neural parametric singing synthesizer modeling
  timbre and expression from natural songs,'' \emph{Applied Sciences}, vol.~7,
  no.~12, p. 1313, 2017.

\bibitem{chandna2019wgansing}
P.~Chandna, M.~Blaauw, J.~Bonada, and E.~Gomez, ``Wgansing: A multi-voice
  singing voice synthesizer based on the wasserstein-gan.''\hskip 1em plus
  0.5em minus 0.4em\relax IEEE, 2019, pp. 1--5.

\bibitem{lee2019adversarially}
J.~Lee, H.-S. Choi, C.-B. Jeon, J.~Koo, and K.~Lee, ``Adversarially trained
  end-to-end korean singing voice synthesis system,'' 2019.

\bibitem{polyak2020ucdsvc}
A.~Polyak, L.~Wolf, Y.~Adi, and Y.~Taigman, ``{Unsupervised Cross-Domain
  Singing Voice Conversion},'' in \emph{Proc. Interspeech 2020}, 2020, pp.
  801--805.

\bibitem{valle2020mellotron}
R.~Valle, J.~Li, R.~Prenger, and B.~Catanzaro, ``Mellotron: Multispeaker
  expressive voice synthesis by conditioning on rhythm, pitch and global style
  tokens,'' in \emph{ICASSP 2020-2020 IEEE International Conference on
  Acoustics, Speech and Signal Processing (ICASSP)}.\hskip 1em plus 0.5em minus
  0.4em\relax IEEE, 2020, pp. 6189--6193.

\bibitem{shen2018natural}
J.~Shen, R.~Pang, R.~J. Weiss, M.~Schuster, N.~Jaitly, Z.~Yang, Z.~Chen,
  Y.~Zhang, Y.~Wang, R.~Skerrv-Ryan \emph{et~al.}, ``Natural {TTS} synthesis by
  conditioning wavenet on {MEL} spectrogram predictions,'' in \emph{2018 IEEE
  International Conference on Acoustics, Speech and Signal Processing
  (ICASSP)}.\hskip 1em plus 0.5em minus 0.4em\relax IEEE, 2018, pp. 4779--4783.

\bibitem{dhariwal2020jukebox}
P.~Dhariwal, H.~Jun, C.~Payne, J.~W. Kim, A.~Radford, and I.~Sutskever,
  ``Jukebox: A generative model for music,'' \emph{arXiv preprint
  arXiv:2005.00341}, 2020.

\bibitem{angelini2020singing}
O.~Angelini, A.~Moinet, K.~Yanagisawa, and T.~Drugman, ``Singing synthesis:
  with a little help from my attention,'' 2020.

\bibitem{zhang2020duriansc}
L.~Zhang, C.~Yu, H.~Lu, C.~Weng, C.~Zhang, Y.~Wu, X.~Xie, Z.~Li, and D.~Yu,
  ``{DurIAN-SC: Duration Informed Attention Network Based Singing Voice
  Conversion System},'' in \emph{Proc. Interspeech 2020}, 2020, pp. 1231--1235.

\bibitem{ren2020deepsinger}
Y.~Ren, X.~Tan, T.~Qin, J.~Luan, Z.~Zhao, and T.-Y. Liu, ``Deepsinger: Singing
  voice synthesis with data mined from the web,'' in \emph{Proceedings of the
  26th ACM SIGKDD International Conference on Knowledge Discovery \& Data
  Mining}, 2020, pp. 1979--1989.

\bibitem{chen2020hifisinger}
J.~Chen, X.~Tan, J.~Luan, T.~Qin, and T.-Y. Liu, ``Hifisinger: Towards
  high-fidelity neural singing voice synthesis,'' 2020.

\bibitem{ren2019fastspeech}
Y.~Ren, Y.~Ruan, X.~Tan, T.~Qin, S.~Zhao, Z.~Zhao, and T.-Y. Liu, ``Fastspeech:
  Fast, robust and controllable text to speech,'' \emph{arXiv preprint
  arXiv:1905.09263}, 2019.

\bibitem{gu2021bytesing}
Y.~Gu, X.~Yin, Y.~Rao, Y.~Wan, B.~Tang, Y.~Zhang, J.~Chen, Y.~Wang, and Z.~Ma,
  ``Bytesing: A chinese singing voice synthesis system using duration allocated
  encoder-decoder acoustic models and wavernn vocoders,'' 2021.

\bibitem{prosodycontrol2}
M.~Christidou, A.~Vioni, N.~Ellinas, G.~Vamvoukakis, K.~Markopoulos,
  P.~Kakoulidis, J.~S. Sung, H.~Park, A.~Chalamandaris, and P.~Tsiakoulis,
  ``Improved prosodic clustering for multispeaker and
  speaker-independentphoneme-level prosody control,'' in \emph{SPECOM}, 2021
  [SUBMITTED].

\bibitem{charpentier1986diphone}
F.~Charpentier and M.~Stella, ``Diphone synthesis using an overlap-add
  technique for speech waveforms concatenation,'' in \emph{ICASSP'86. IEEE
  International Conference on Acoustics, Speech, and Signal Processing},
  vol.~11.\hskip 1em plus 0.5em minus 0.4em\relax IEEE, 1986, pp. 2015--2018.

\bibitem{verhelst1993overlap}
W.~Verhelst and M.~Roelands, ``An overlap-add technique based on waveform
  similarity (wsola) for high quality time-scale modification of speech,'' in
  \emph{1993 IEEE International Conference on Acoustics, Speech, and Signal
  Processing}, vol.~2.\hskip 1em plus 0.5em minus 0.4em\relax IEEE, 1993, pp.
  554--557.

\bibitem{raptis2016expressive}
S.~Raptis, P.~Tsiakoulis, A.~Chalamandaris, and S.~Karabetsos, ``Expressive
  speech synthesis for storytelling: the innoetics’ entry to the blizzard
  challenge 2016,'' in \emph{Proc. Blizzard Challenge}, 2016.

\bibitem{boersma2006praat}
P.~Boersma, ``Praat: doing phonetics by computer,'' \emph{http://www. praat.
  org}, 2006.

\bibitem{prosodycontrol}
A.~Vioni, M.~Christidou, N.~Ellinas, G.~Vamvoukakis, P.~Kakoulidis, T.~Kim,
  J.~S. Sung, H.~Park, A.~Chalamandaris, and P.~Tsiakoulis, ``Prosodic
  clustering for phoneme-level prosody control in end-to-end speech
  synthesis,'' in \emph{Proc. ICASSP}, 2021.

\bibitem{lpctron}
N.~Ellinas, G.~Vamvoukakis, K.~Markopoulos, A.~Chalamandaris, G.~Maniati,
  P.~Kakoulidis, S.~Raptis, J.~S. Sung, H.~Park, and P.~Tsiakoulis, ``High
  quality streaming speech synthesis with low, sentence-length-independent
  latency,'' in \emph{Proc. Interspeech}, 2020.

\bibitem{Zhang2019}
\BIBentryALTinterwordspacing
Y.~Zhang, R.~J. Weiss, H.~Zen, Y.~Wu, Z.~Chen, R.~Skerry-Ryan, Y.~Jia,
  A.~Rosenberg, and B.~Ramabhadran, ``{Learning to Speak Fluently in a Foreign
  Language: Multilingual Speech Synthesis and Cross-Language Voice Cloning},''
  in \emph{Proc. Interspeech 2019}, 2019, pp. 2080--2084. [Online]. Available:
  \url{http://dx.doi.org/10.21437/Interspeech.2019-2668}
\BIBentrySTDinterwordspacing

\bibitem{laprie1998automatic}
Y.~Laprie and V.~Colotte, ``Automatic pitch marking for speech transformations
  via td-psola,'' in \emph{9th European Signal Processing Conference (EUSIPCO
  1998)}.\hskip 1em plus 0.5em minus 0.4em\relax IEEE, 1998, pp. 1--4.

\bibitem{Chalamandaris2009}
A.~Chalamandaris, P.~Tsiakoulis, S.~Raptis, and S.~Karabetsos, ``{Corpus design
  for a unit selection TTS system with application to Bulgarian},'' in
  \emph{Proc. 4th Conference on Human language technology: challenges for
  computer science and linguistics}, 2009, pp. 35--46.

\bibitem{vipperla2020bunched}
R.~Vipperla, S.~Park, K.~Choo, S.~Ishtiaq, K.~Min, S.~Bhattacharya,
  A.~Mehrotra, A.~G.~C. Ramos, and N.~D. Lane, ``Bunched lpcnet: Vocoder for
  low-cost neural text-to-speech systems,'' \emph{arXiv preprint
  arXiv:2008.04574}, 2020.

\bibitem{srivastava2014dropout}
N.~Srivastava, G.~Hinton, A.~Krizhevsky, I.~Sutskever, and R.~Salakhutdinov,
  ``Dropout: a simple way to prevent neural networks from overfitting,''
  \emph{The journal of machine learning research}, vol.~15, no.~1, pp.
  1929--1958, 2014.

\bibitem{krueger2016zoneout}
D.~Krueger, T.~Maharaj, J.~Kram{\'a}r, M.~Pezeshki, N.~Ballas, N.~R. Ke,
  A.~Goyal, Y.~Bengio, A.~Courville, and C.~Pal, ``Zoneout: Regularizing rnns
  by randomly preserving hidden activations,'' \emph{arXiv preprint
  arXiv:1606.01305}, 2016.

\bibitem{kingma2015adam}
D.~P. Kingma and J.~Ba, ``Adam: A methodfor stochastic optimization,'' in
  \emph{International Conference onLearning Representations (ICLR)}, 2015.

\end{thebibliography}


\end{document}